\documentclass[pdflatex,sn-mathphys-num]{sn-jnl}

\usepackage{graphicx}%
\usepackage{multirow}%
\usepackage{amsmath,amssymb,amsfonts}%
\usepackage{amsthm}%
\usepackage{mathrsfs}%
\usepackage[title]{appendix}%
\usepackage{xcolor}%
\usepackage{textcomp}%
\usepackage{manyfoot}%
\usepackage{booktabs}%
\usepackage{algorithm}%
\usepackage{algorithmicx}%
\usepackage{algpseudocode}%
\usepackage{listings}%
\usepackage{subcaption} 
\usepackage{longtable}

\theoremstyle{thmstyleone}%
\theoremstyle{thmstyletwo}%

\theoremstyle{thmstylethree}%

\begin{document}

\title[Article Title]{Integration of Wavelet Transform Convolution and Channel Attention with LSTM for Stock Price Prediction based Portfolio Allocation}


\author*[1]{\fnm{Junjie} 
\sur{Guo}}\email{jg1806@scarletmail.rutgers.edu}

\affil*[1]{\orgdiv{Rutgers Business School}, \orgname{Rutgers University}, \orgaddress{\street{1 Washington Pl}, \city{Newark}, \postcode{07102}, \state{New Jersey}, \country{USA}}}

\abstract{Portfolio allocation via stock price prediction is inherently difficult due to the notoriously low signal-to-noise ratio of stock time series. This paper proposes a method by integrating wavelet transform convolution and channel attention with LSTM to implement stock price prediction based portfolio allocation. Stock time series data first are processed by wavelet transform convolution to reduce the noise. Processed features are then reconstructed by channel attention. LSTM is utilized to predict the stock price using the final processed features. We construct a portfolio consists of four stocks with trading signals predicted by model. Experiments are conducted by evaluating the return, Sharpe ratio and max drawdown performance. The results indicate that our method achieves robust performance even during period of post-pandemic downward market.}




\maketitle

\section{Introduction}\label{sec1}

Machine learning algorithms, exemplified by deep neural networks, have achieved attractive accomplishments across many industries. Benefited from this advancement, financial industry now empower its with deep learning models\cite{bib1}. Quantitative trading is one of the many fields that rely on predictive models, has turned its attention from linear models to deep neural networks driven non-linear models.

Portfolio allocation problem sits at the core of finance not only because its importance in real-world applications but because its intrinsically difficulty to handle with. Mean-Variance portfolio optimization takes both return and risk into account which is a non-parametric approach to estimate the weights of assets in a portfolio\cite{bib2}. Though as an explainable tool, its simplicity can not capture the movement complex non-linear system especially when it comes to dynamically rebalanced portfolios\cite{bib3}\cite{bib4}.

Stock price prediction based portfolio allocation is one specific type of approach which utilize a wide range of models from linear models to non-linear models. Stock price prediction aims to predict the future stock price using previous features of stock\cite{bib5}. Portfolios then can be constructed by longing stocks with positive price increment and shorting stocks with negative price increment\cite{bib6}. This approach of portfolio allocation heavily rely on the model capability to capture the dynamically evolving movement of stock price.

Linear models assume the linearity of the system that naturally neglect the complex dynamics of stock systems\cite{bib7}. Therefore, linear models applications in stock price prediction can be limited due to its drawbacks. In comparison, deep learning models which compose of neural networks, are non-linear predictors\cite{bib8}\cite{bib9}\cite{bib10}. Deep learning models advantages over traditional linear models made itself a favorable in stock price prediction problem. 

Though with many powerful models in price prediction, the inherently low signal-to-ratio problem of stock time series remains a major challenge hindering its application in portfolio allocation\cite{bib13}\cite{bib14}. The failure of identifying non-linear dynamics due to noisy data may lead to non-robust trading signals construction and underperform the portfolio. 

This paper propose an integrated approach taking the existing problems into account and mitigate the effect of those long-standing problems. Specially, we use wavelet transform to denoise the original stock features and transform those features by channel attention\cite{bib11}\cite{bib12}. Those final features are as input of the LSTM predictor to predict the one step ahead stock price\cite{bib17}. A trading portfolio is constructed using four fixed stocks with equal weights. The long and short positions are determined by the price prediction, positive and negative price increment indicate long position and short position. The portfolio is dynamically rebalanced by assigning different stocks to long and short positions. We evaluate the portfolio performance by return, Sharpe ratio and maximum drawdown.

\section{Related work}\label{sec2}

Traditional stock price time series prediction methods have primarily relied on statistical models such as Autoregressive Integrated Moving Average (ARIMA) and Generalized Autoregressive Conditional Heteroskedasticity (GARCH)\cite{bib18}\cite{bib19}. These models aim to capture linear dependencies and volatility clustering in financial data, providing a foundation for understanding stock price movements.

With advancements in computational capabilities, machine learning techniques have been employed to enhance stock price predictions by modeling nonlinear patterns. Support Vector Machines (SVM), for instance, have been utilized to forecast financial time series, demonstrating improved accuracy over traditional statistical methods\cite{bib20}\cite{bib21}. These methods leverage historical data to find complex relationships that may not be apparent through conventional analysis.

Deep learning approaches have further revolutionized stock price prediction by employing neural networks capable of learning hierarchical feature representations\cite{bib20}\cite{bib21}. Long Short-Term Memory (LSTM) networks, a type of recurrent neural network, have been particularly effective in capturing temporal dependencies in financial data. Fischer and Krauss (2018) showed that LSTM models outperform traditional machine learning methods in predicting stock returns\cite{bib22}.

In portfolio allocation, both traditional methods like the Markowitz Mean-Variance Optimization and deep learning techniques have been explored to optimize investment strategies\cite{bib23}\cite{bib24}. Deep reinforcement learning frameworks have been proposed to adaptively allocate assets, learning optimal policies through interactions with the market environment\cite{bib25}. Jiang et al. (2017) demonstrated that such frameworks could effectively manage portfolios by maximizing returns while controlling risk\cite{bib26}.

To address the noise inherent in stock time series data, deep learning methods incorporating denoising techniques have been developed\cite{bib27}\cite{bib28}. Autoencoders, particularly stacked denoising autoencoders, have been employed to extract robust features from noisy inputs, enhancing prediction performance. Bao et al. (2017) integrated wavelet transforms with LSTM networks to reduce noise and improve the accuracy of financial time series forecasting\cite{bib28}.

Wavelet transform methods have been applied to time series prediction to analyze data across multiple scales and frequencies\cite{bib29}. By decomposing time series into different frequency components, wavelet transforms help in capturing both local and global patterns. Zhang et al. (2017) utilized multi-frequency trading patterns discovered through wavelet transforms to predict stock prices effectively\cite{bib30}.

Channel attention mechanisms have recently gained attention in time series prediction for their ability to focus on informative features across different channels\cite{bib32}. By assigning dynamic weights to feature channels, models can enhance relevant signals while suppressing noise. Woo et al. (2018) introduced the Convolutional Block Attention Module (CBAM), which incorporates channel attention and has been applied to improve model performance in various domains, including time series analysis\cite{bib31}.

\section{Preliminary}\label{sec3}

In this section, we provide the theoretical background pertinent to our study, including stock price prediction via deep learning, long-short stock portfolio and evaluation metrics of portfolio.

\subsection{Stock Price Prediction via Deep Learning}

In the context of stock price prediction, deep learning models learn patterns from historical data to forecast future price. This paper studies one-step ahead stock price prediction.

In supervised deep learning, the objective is to learn a function \( f_{\theta} \) that maps input data \( \mathbf{x} \) to output targets \( \mathbf{y} \), based on a set of labeled training examples. The function \( f_{\theta} \) is parameterized by a set of parameters \( \theta \), which are optimized during training.

Given a training dataset \( \mathcal{D} = \{ (\mathbf{x}^{(i)}, \mathbf{y}^{(i)}) \}_{i=1}^{N} \), where \( \mathbf{x}^{(i)} \in \mathbb{R}^{n} \) is the input feature vector, \( \mathbf{y}^{(i)} \in \mathbb{R}^{m} \) is the corresponding target output. A deep learning model can be described as a mapping  \( f_{\theta}: \mathbb{R}^{n} \rightarrow \mathbb{R}^{m} \), parameterized by \( \theta \). In this formulation, let \( \mathbf{X}_t \) represent the feature vector at time \( t \), the objective is to predict the next day's price \( r_{t+1} \):

\[ \hat{p}_{t+1} = f_{\theta}(\mathbf{X}_t) \]

\subsection{Long-short Stock Portfolio}
A long-short stock portfolio is a collection of stock assets that contains stocks with long positions and short positions. Those stocks with short positions can be denoted as a set: \(\mathcal{S} = \{s_i \mid i=1,...,N_1\} \). Similarly, stocks with positive positions can be denoted as a set: \(\mathcal{L} = \{l_t \mid t=1,...,N_2\} \). The union of the two set \(\mathcal{P_{LS}} = \{ s_i \mid i=1,...,N_1 \} \bigcup  \{l_t \mid t=1,...,N_2\} \) is called the long-short portfolio. The amount of stocks in the long-short portfolio is \( N_1 + N_2\).

In stock price prediction based portfolio allocation, at each time step the portfolio is rebalanced by assigning stocks as long or short positions. We study the portfolio with fixed stocks and rebalance frequency of one day. 

\subsection{Portfolio Evaluation Metrics}

Return of a portfolio is a indicator measures its profitability which directly reflect the quality of the performance. We consider the return at each trading day and return over the whole trading period.

Assume we have \(N \) stocks in the long-short portfolio. For each stock \(S_i\) we can obtain its price prediction over a T-day trading period. We denote the prediction result as a set for each stock: \(\hat{S_i} = \{\hat{p}^{i}_{t} \mid t =1,...,T \}\). The corresponding true price set can be expressed as \({S_i} = \{{p}^{i}_{t} \mid t =1,...,T \}\). The true return of stock i can be denoted as \({R_i} = \{{r}^{i}_{t} \mid t =1,...,T \}\). To calculate the return of the portfolio we need to determine the weights of stocks at each time step. This study uses equal weight for each stock at each trading day. The weights \(\mathcal{W}_p \) can be determined as:

\[
\mathcal{W}_p = \frac{\mathcal{J}}{N} = 
\begin{bmatrix} 
{1}/{N} \\ 
{1}/{N} \\
\vdots \\
{1}/{N}
\end{bmatrix}
\]

Where \(\mathcal{J}\) is a \(N \) dimensional vector with all elements equal to 1. \(N \) is the number of stocks in the portfolio.  \(\mathcal{W}_p\) has the same dimension as \(\mathcal{J}\) which is \(N\).

We define an indicator function which maps the input to a binary output. If the difference of price at time \(t\) and \(t-1\) is less than 0, we assign the output as -1. If the difference of price at time \(t\) and \(t-1\) is greater than 0, we assign the output as 1. The indicator function can be denoted as:

\[
\mathcal{I}(\hat{p}_{t}) = \begin{cases}
    -1, \hat{p}_{t} < {p}_{t-1} \\
    1, \hat{p}_{t} > {p}_{t-1}
\end{cases}
\]

The portfolio return \({d}_{r}\) at each trading day is the summation of returns of long and short positions. This is can be calculated as:

\[
d_{r} = \sum^{N}_{i=1} \mathcal{I}(\hat{p}^{t}_{i}) {w}^{t}_{i} {R}^{t}_{i}
\]

The portfolio total return \({t}_{r}\) over a T-day period then can be calculated as: 

\[
{t}_{r} = \prod^{T}_{t=1}(\sum^{N}_{i=1} \mathcal{I}(\hat{p}^{t}_{i}) {w}^{t}_{i} {R}^{t}_{i} + 1 ) -1 
\]

To more precisely assess the portfolio taking risk into account, Sharpe ratio can be leveraged as a metric to indicate the risk-adjusted return of the portfolio. High Sharpe ratio is considered better than low Sharpe ratio. Sharpe ratio is defined as:

\[
\text{Sharpe Ratio} = \frac{E[R_p] - R_f}{\sigma_p}\
\]

where \( E[R_p] \) is the expected portfolio return, \( R_f \) is the risk-free rate, \( \sigma_p \) is the standard deviation of portfolio returns.

Max Drawdown (MDD) is a risk metric used to assess the largest peak-to-trough decline in the value of an investment portfolio, typically over a specified time period. It represents the most significant loss from the highest value (peak) to the lowest value (trough) before a new peak is achieved. MDD is commonly used to measure the worst-case scenario for an investor in terms of how much value they could lose during a period of market downturns.

Let \( V_t \) be the cumulative portfolio value at time \( t \). The drawdown at time \( t \) is:
\[
\text{Drawdown}_t = \frac{V_t - \max\limits_{s \leq t} V_s}{\max\limits_{s \leq t} V_s}\
\]

The maximum drawdown defined on the T time steps trading period is:
\[
\text{Maximum Drawdown} = \left| \min\limits_{t \in [1, T]} \left( \text{Drawdown}_t \right) \right|
\]

\section{Methodology}\label{sec4}

In this section, we present a novel approach for time series prediction, which integrates signal processing techniques with deep learning models to enhance feature extraction and improve predictive accuracy. Specifically, we combine the Discrete Cosine Transform (DCT), Wavelet Transform (WT), and Long Short-Term Memory (LSTM) networks into a unified model. The goal is to capture both frequency-domain and temporal dependencies in the data, enabling better performance in forecasting tasks.

\subsection{Feature Extraction with Wavelet Transform (WT)}

In addition to the frequency-domain features extracted via DCT, we apply a wavelet transform to capture both time and frequency information. The Discrete Wavelet Transform (DWT) decomposes the input signal into approximation and detail coefficients, which are then processed using convolutional layers.

The forward DWT for a signal \( \mathbf{x}_n \) is expressed as:

\[
a_{j,k} = \sum_{n} x_n \phi_{j,k}(n)
\]
\[ 
d_{j,k} = \sum_{n} x_n \psi_{j,k}(n)
\]

where \( \phi_{j,k} \) and \( \psi_{j,k} \) are the scaling and wavelet functions, respectively, at scale \( j \) and position \( k \). The approximation coefficients \( a_{j,k} \) capture the low-frequency components, while the detail coefficients \( d_{j,k} \) capture high-frequency components.

To reduce the complexity of the wavelet transform, we use depthwise separable convolutions, which decompose the convolution process into two stages: a depthwise convolution applied to each input channel separately and a pointwise convolution to combine the outputs across channels.

The Wavelet Transform Convolutional Layer (WTConv1d) operates as follows:

\begin{enumerate}
    \item Perform wavelet decomposition using wavelet filters:
    \[
    [\mathbf{C}_{\text{low}}, \mathbf{C}_{\text{high}}] = \text{WT}(\mathbf{X}, \mathbf{F}_{\text{dec}})
    \]
    where \( \mathbf{F}_{\text{dec}} \) are the decomposition filters.
    \item Apply a convolution operation to the decomposed coefficients:
    \[
    \mathbf{C}' = \gamma \cdot \left( \mathbf{W}_{\text{wavelet}} \ast [\mathbf{C}_{\text{low}}, \mathbf{C}_{\text{high}}] \right)
    \]
    where \( \gamma \) is a learnable scaling parameter and \( \ast \) denotes convolution.
    \item Reconstruct the signal from the wavelet coefficients:
    \[
    \mathbf{X}' = \text{IWT}(\mathbf{C}', \mathbf{F}_{\text{rec}})
    \]
    where \( \mathbf{F}_{\text{rec}} \) are the reconstruction filters.
\end{enumerate}

\subsection{Preprocessing with Discrete Cosine Transform (DCT)}

The Discrete Cosine Transform (DCT) is applied to the input time series data to extract frequency-domain features, which are known to compact the signal energy effectively. The DCT is applied to each input sequence to identify the most significant frequency components.

For a given input sequence \( \mathbf{x}_n \) of length \( N \), the DCT is computed as:

\[
X_k = \sum_{n=0}^{N-1} x_n \cos\left( \frac{\pi}{N} \left( n + \frac{1}{2} \right) k \right), \quad k = 0, 1, \dots, N-1
\]

where \( X_k \) represents the DCT coefficients at frequency \( k \), and \( x_n \) is the input value at time step \( n \). The DCT captures both low-frequency and high-frequency components of the time series, allowing the model to focus on the most relevant features for prediction.

The DCT is computed efficiently using the Fast Fourier Transform (FFT) algorithm. Following the DCT computation, a channel-wise attention mechanism is applied to emphasize the most informative frequency components.

\subsection{Channel Attention via DCT Coefficients}

To enhance the model's ability to focus on the most significant features, we introduce a channel attention mechanism that operates on the DCT coefficients. The channel attention mechanism assigns higher weights to important channels, thereby improving feature extraction. The attention weights \( \mathbf{W}_c \) are computed using a two-layer fully connected neural network applied to the DCT-transformed input.

The channel-wise attention mechanism is defined as follows:

\begin{enumerate}
    \item Apply the DCT to each channel of the input \( \mathbf{X}_{:,c,:} \) for \( c \in [1, C] \), where \( C \) is the number of channels:
    \[
    \mathbf{F}_c = \text{DCT}(\mathbf{X}_{:,c,:}).
    \]
    \item Stack the DCT-transformed features across all channels:
    \[
    \mathbf{F} = \text{Stack}(\mathbf{F}_1, \dots, \mathbf{F}_C) \in \mathbb{R}^{B \times C \times L}
    \]
    where \( B \) is the batch size, and \( L \) is the sequence length.
    \item Apply layer normalization to the DCT-transformed features:
    \[
    \hat{\mathbf{F}} = \text{LayerNorm}(\mathbf{F}).
    \]
    \item Compute the attention weights \( \mathbf{W} \) using a fully connected neural network:
    \[
    \mathbf{W} = \sigma\left( \mathbf{W}_2 \cdot \text{ReLU}(\mathbf{W}_1 \hat{\mathbf{F}} + \mathbf{b}_1) + \mathbf{b}_2 \right)
    \]
    where \( \sigma \) denotes the sigmoid activation function, and \( \mathbf{W}_1 \in \mathbb{R}^{C \times 2C} \), \( \mathbf{W}_2 \in \mathbb{R}^{2C \times C} \), \( \mathbf{b}_1 \) and \( \mathbf{b}_2 \) are bias vectors.
    \item The final output of the channel attention mechanism is obtained by element-wise multiplication of the input with the attention weights:
    \[
    \mathbf{Y} = \mathbf{X} \odot \mathbf{W}
    \]
    where \( \odot \) denotes element-wise multiplication.
\end{enumerate}

\subsection{Temporal Feature Modeling with LSTM}

The Long Short-Term Memory (LSTM) network is employed to model the temporal dependencies in the time series data. LSTM units consist of input, forget, and output gates, which allow the network to learn long-term dependencies and mitigate the vanishing gradient problem.

The LSTM update equations are as follows:

\[
\mathbf{i}_t = \sigma(\mathbf{W}_i \mathbf{x}_t + \mathbf{U}_i \mathbf{h}_{t-1} + \mathbf{b}_i),
\]
\[
\mathbf{f}_t = \sigma(\mathbf{W}_f \mathbf{x}_t + \mathbf{U}_f \mathbf{h}_{t-1} + \mathbf{b}_f),
\]
\[
\mathbf{o}_t = \sigma(\mathbf{W}_o \mathbf{x}_t + \mathbf{U}_o \mathbf{h}_{t-1} + \mathbf{b}_o),
\]
\[
\mathbf{g}_t = \tanh(\mathbf{W}_g \mathbf{x}_t + \mathbf{U}_g \mathbf{h}_{t-1} + \mathbf{b}_g),
\]
\[
\mathbf{c}_t = \mathbf{f}_t \odot \mathbf{c}_{t-1} + \mathbf{i}_t \odot \mathbf{g}_t,
\]
\[
\mathbf{h}_t = \mathbf{o}_t \odot \tanh(\mathbf{c}_t),
\]

where \( \sigma \) is the sigmoid function, \( \tanh \) is the hyperbolic tangent, and \( \odot \) denotes element-wise multiplication. The hidden state \( \mathbf{h}_t \) is updated at each time step, capturing the temporal dynamics of the input data.

\subsection{Final Prediction}

The final prediction is obtained by passing the LSTM outputs through a linear layer:

\[
\mathbf{Y} = \mathbf{W}_{\text{out}} \mathbf{H}_T + \mathbf{b}_{\text{out}},
\]

where \( \mathbf{H}_T \) is the hidden state at the last time step, \( \mathbf{W}_{\text{out}} \) is the output weight matrix, and \( \mathbf{b}_{\text{out}} \) is the output bias.

\subsection{Model Training and Optimization}

The model is trained using the Adam optimizer with a learning rate schedule to adaptively adjust the learning rate during training. The Mean Squared Error (MSE) loss function is employed for regression tasks, defined as:

\[
\mathcal{L} = \frac{1}{N} \sum_{i=1}^{N} (\hat{y}_i - y_i)^2,
\]

where \( \hat{y}_i \) is the predicted value and \( y_i \) is the ground truth value.

\section{Experiment and Result}\label{sec5}

In this section, we present a detailed account of our experimental setup and the corresponding results obtained from our proposed methodology. The experiments are structured into several subsections covering data collection, benchmark strategy formulation, price prediction, individual stock trading returns, portfolio trading, overall performance evaluation, and the assessment of prediction metrics.

\subsection{Data Collection}

The empirical analysis is based on historical daily data for four major stocks: AAPL, AMZN, GE, and MSFT. The data spans from October 1, 2013, to September 29, 2023, providing a comprehensive view of market behavior over nearly a decade. The dataset is partitioned into two subsets: 80\% is allocated for training our predictive model, and the remaining 20\% is reserved for testing. 

\begin{figure}[!h]
    \begin{subfigure}{0.5\textwidth}
        \includegraphics[width=\linewidth]{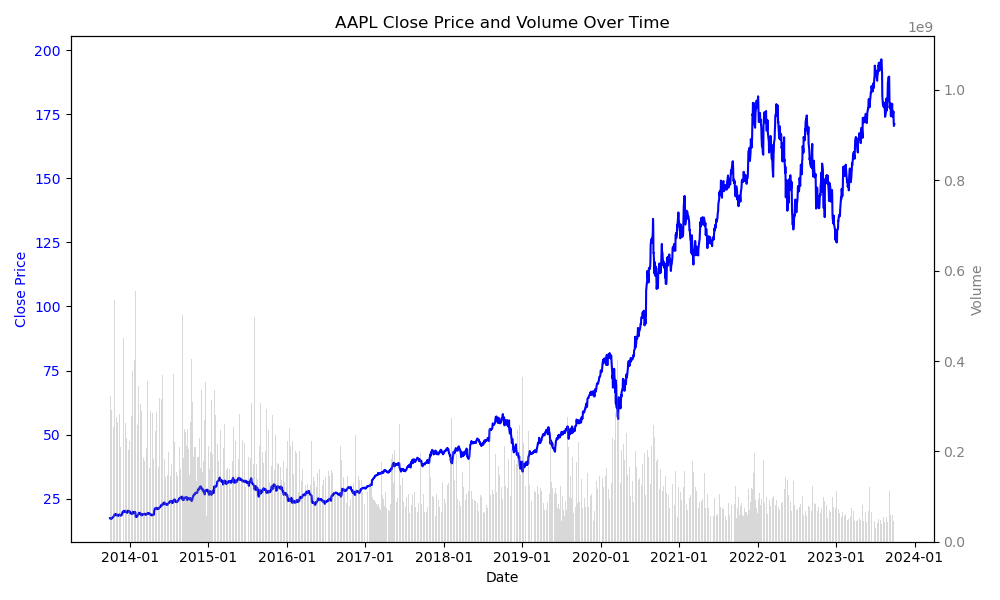}
        \caption{}
    \end{subfigure}
    \hfill
    \begin{subfigure}{0.5\textwidth}
        \includegraphics[width=\linewidth]{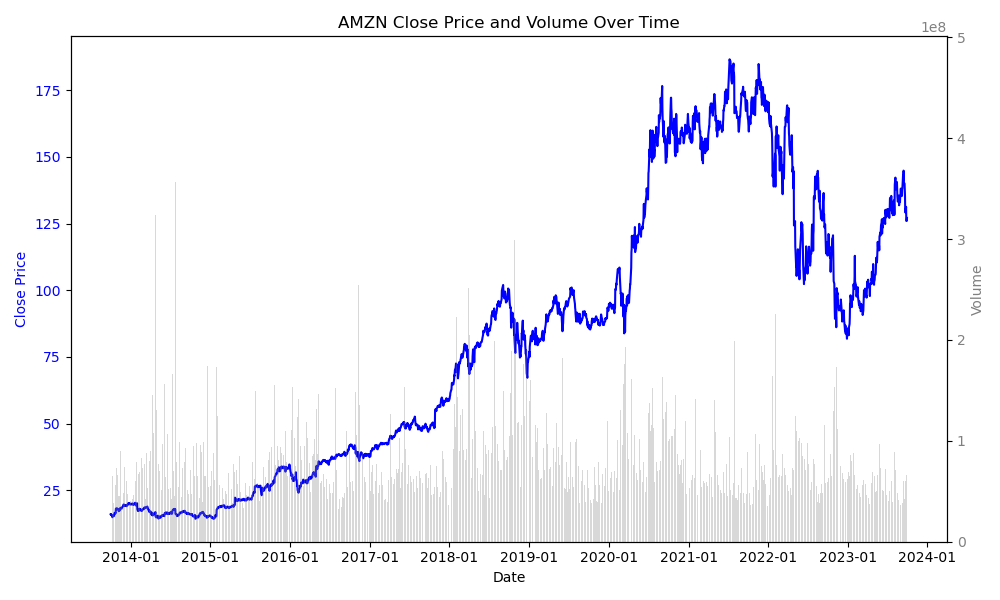}
        \caption{}
    \end{subfigure}
    \hfill
    \begin{subfigure}{0.5\textwidth}
        \includegraphics[width=\linewidth]{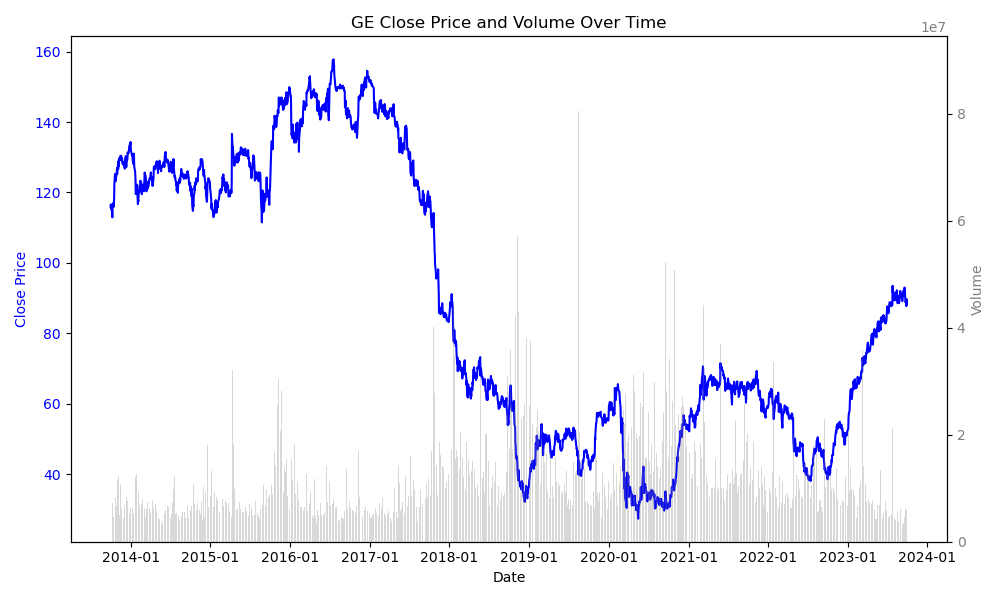}
        \caption{}
    \end{subfigure}
    \hfill
    \begin{subfigure}{0.5\textwidth}
        \includegraphics[width=\linewidth]{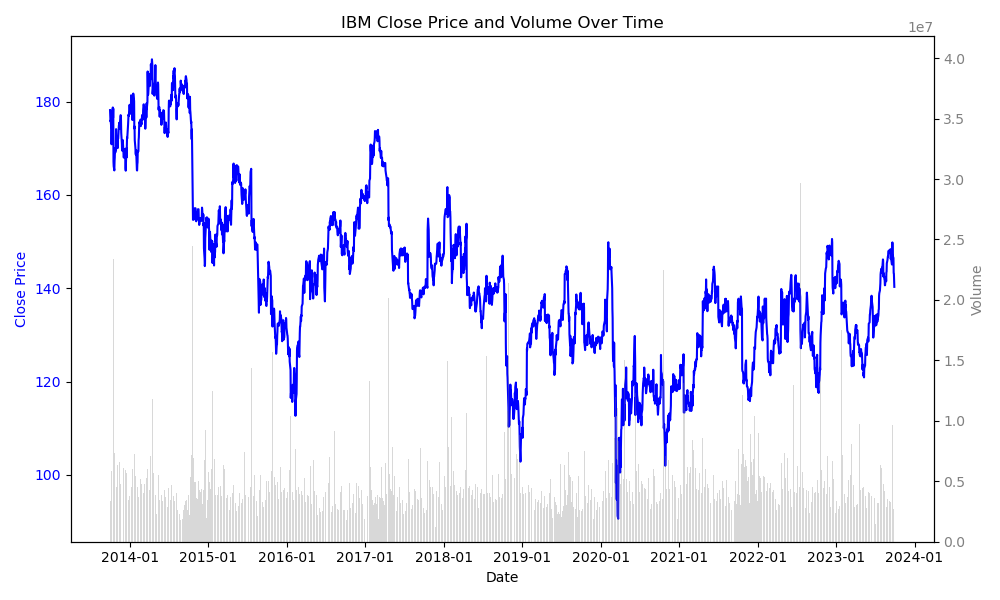}
        \caption{}
    \end{subfigure}
    \hfill
    \caption{Selected S\&P 500 stocks with prices and volumes}
    \label{fig:multi_figs}
\end{figure}
The features extracted for model training include open, high, low, volume, and close prices. These features have been chosen because of their relevance in capturing intrinsic market dynamics and volatility patterns. Figures. 1 illustrates the price and volume trajectories for each of the selected stocks, providing a visual insight into their historical performance.

\subsection{Buy and Hold Trading}

The buy and hold strategy is a passive investment approach wherein an investor purchases a security and retains it for an extended period, regardless of fluctuations in the market. This strategy is premised on the notion that, despite short-term volatility, the long-term trend of the market tends to be upward. 

In our study, the buy and hold strategy is employed as a benchmark to evaluate the performance of our proposed trading strategy. The primary motivation for this comparison arises from the post-pandemic market conditions, during which stocks experienced heightened volatility and substantial drawdowns. This environment underscores the limitations of a passive strategy in capturing short-term market opportunities and managing risk.

The return from the buy and hold strategy is calculated using the following formula:

\[
R_{\text{B\&H}} = \frac{P_T - P_0}{P_0},
\]

where \( P_0 \) is the initial stock price and \( P_T \) is the price at the end of the holding period. This metric provides a clear measure of the absolute return that an investor would have realized under a passive buy and hold regime.

\subsection{Price Prediction}

Leveraging our proposed prediction methodology, we obtain price forecasts for the four stocks over the designated backtesting period. The prediction model integrates both technical and fundamental aspects of market behavior, generating not only price forecasts but also actionable trading signals. These signals are classified as long or short, based on the expected price movement.

For comparison, we test the data on other five models(MLP, CNN, LSTM, CNN-LSTM, Attention-LSTM) used widely in stock price prediction task.
\begin{figure}[!h]
    \begin{subfigure}{0.5\textwidth}
        \includegraphics[width=\linewidth]{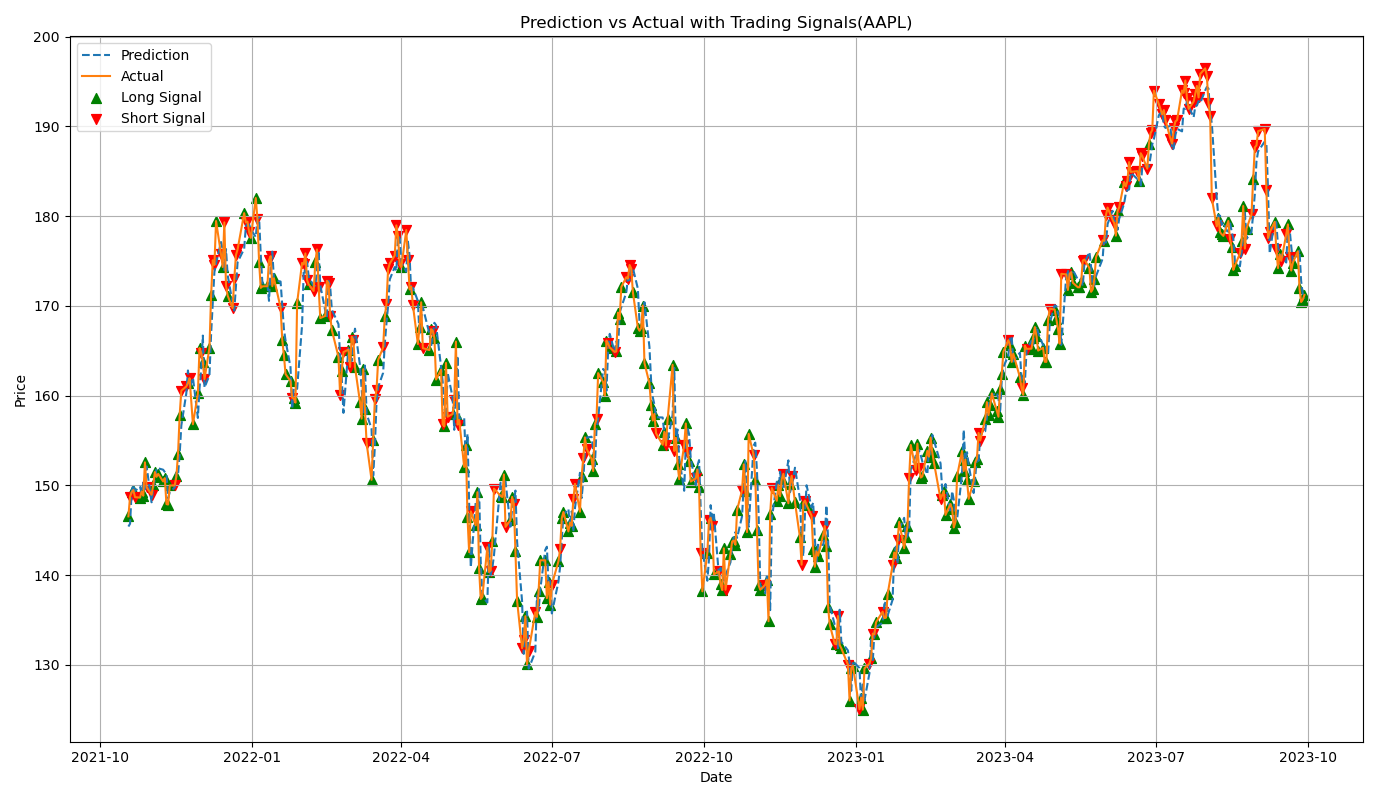}
        \caption{}
    \end{subfigure}
    \hfill
    \begin{subfigure}{0.5\textwidth}
        \includegraphics[width=\linewidth]{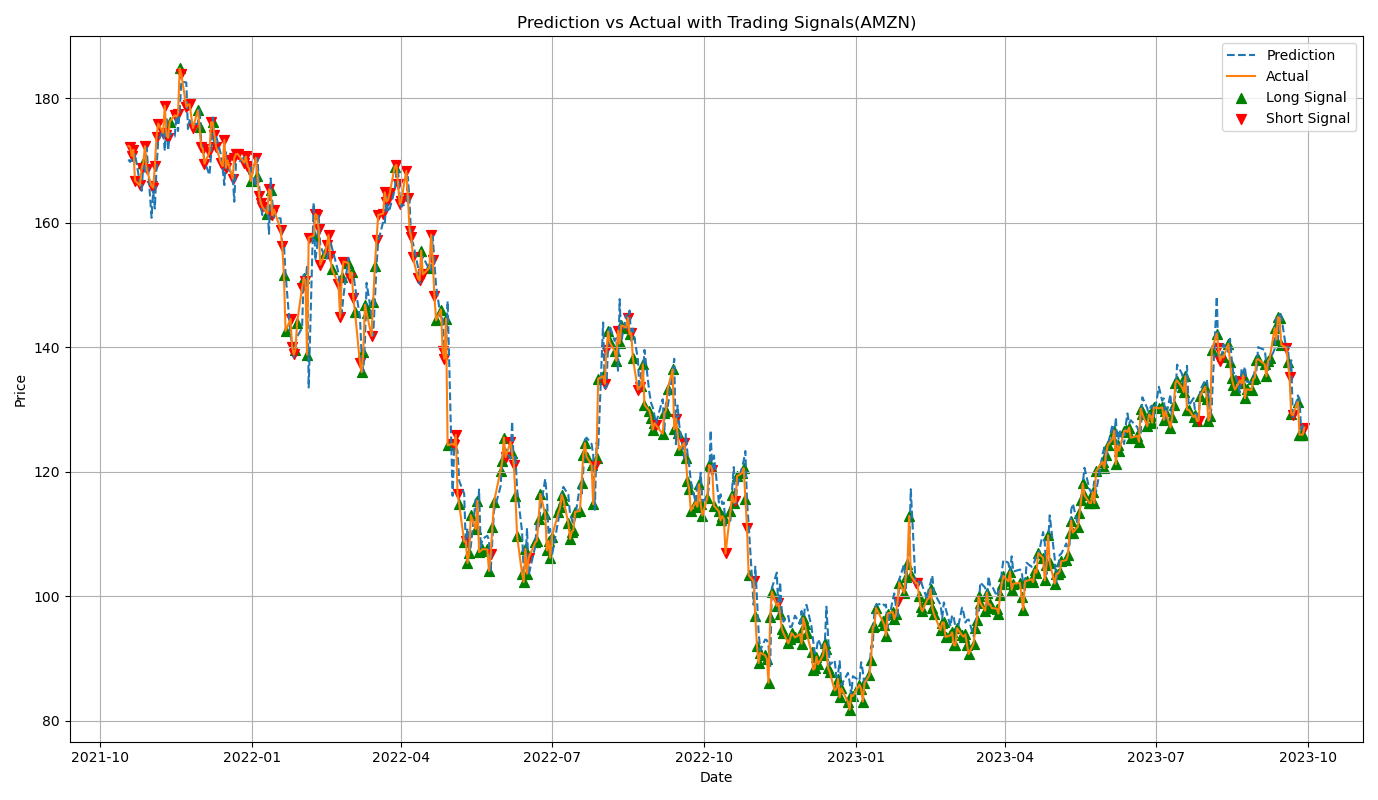}
        \caption{}
    \end{subfigure}
    \hfill
    \begin{subfigure}{0.5\textwidth}
        \includegraphics[width=\linewidth]{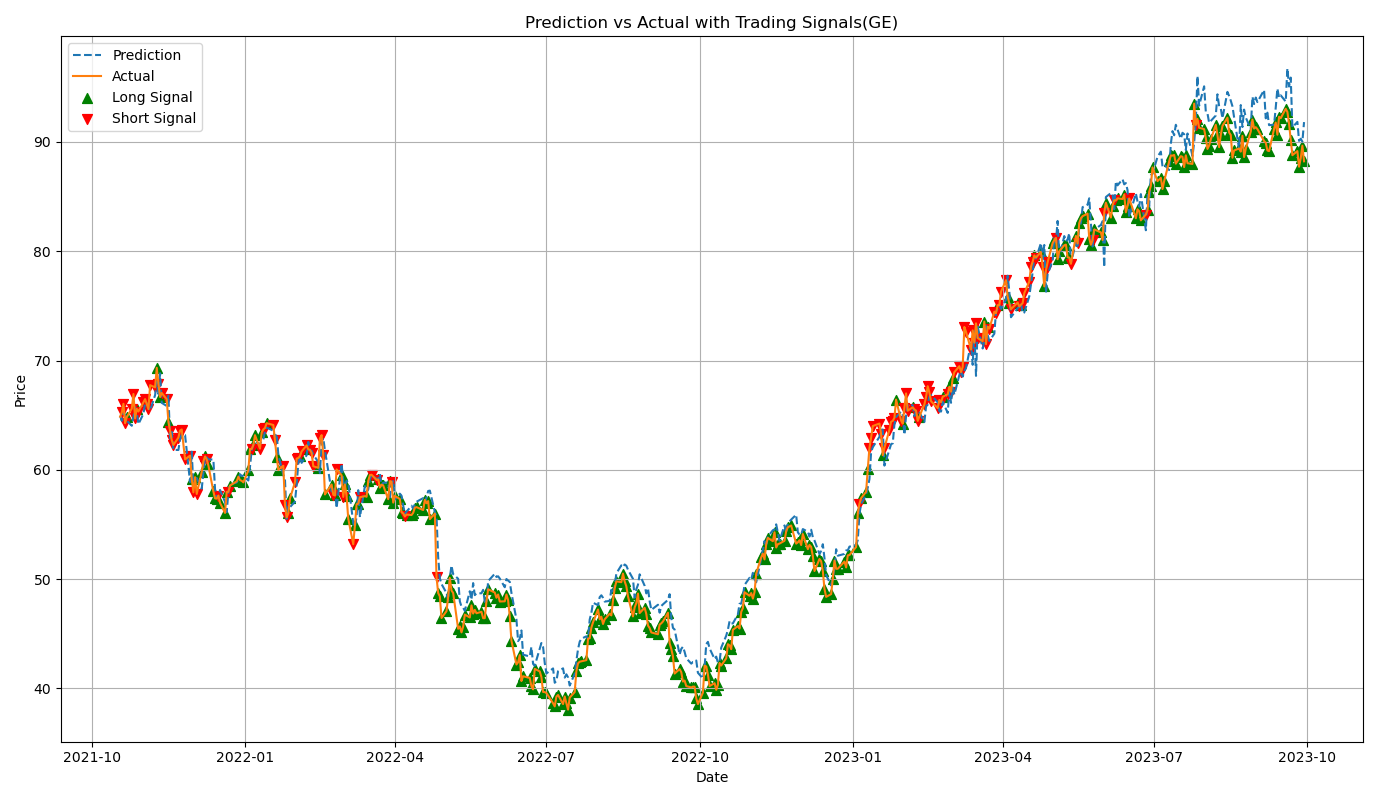}
        \caption{}
    \end{subfigure}
    \hfill
    \begin{subfigure}{0.5\textwidth}
        \includegraphics[width=\linewidth]{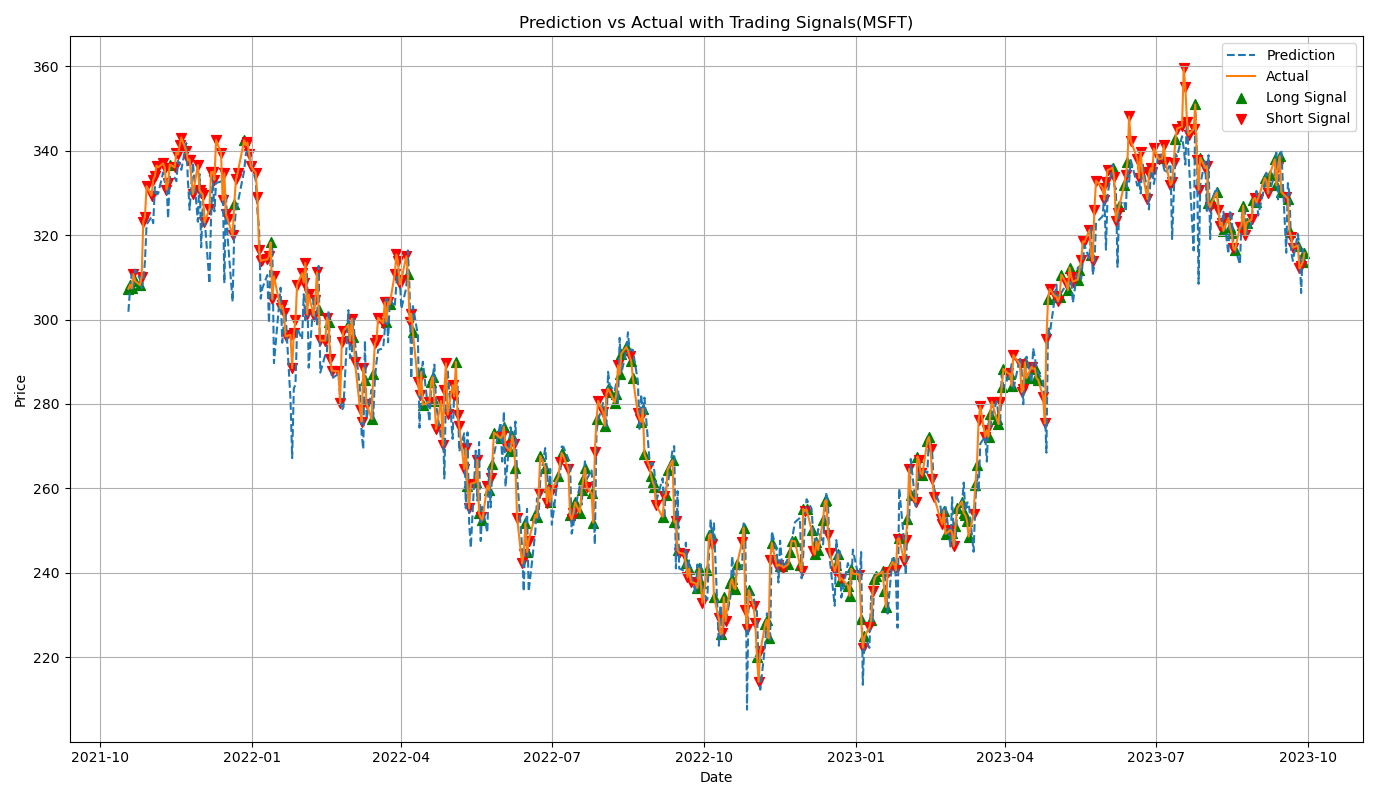}
        \caption{}
    \end{subfigure}
    \hfill
    \caption{Price prediction results with trading signals}
    \label{fig:multi_figs}
\end{figure}

Figures.2 presents the price prediction results for AAPL, AMZN, GE, and MSFT, respectively. Each figure includes annotations for long and short signals, thereby illustrating the model’s capability to identify favorable entry and exit points in the market. The clarity of these figures reinforces the robustness of our predictive framework under volatile market conditions.

\subsection{Trading Returns of Stocks}

Building upon the predicted prices and the corresponding trading signals, we execute individual trading strategies for each stock. In each instance where a trading signal is generated, the entire portfolio allocated to that stock is fully invested. This full-investment approach accentuates the efficacy of our signal generation mechanism.

\begin{figure}[!h]
    \begin{subfigure}{0.5\textwidth}
        \includegraphics[width=\linewidth]{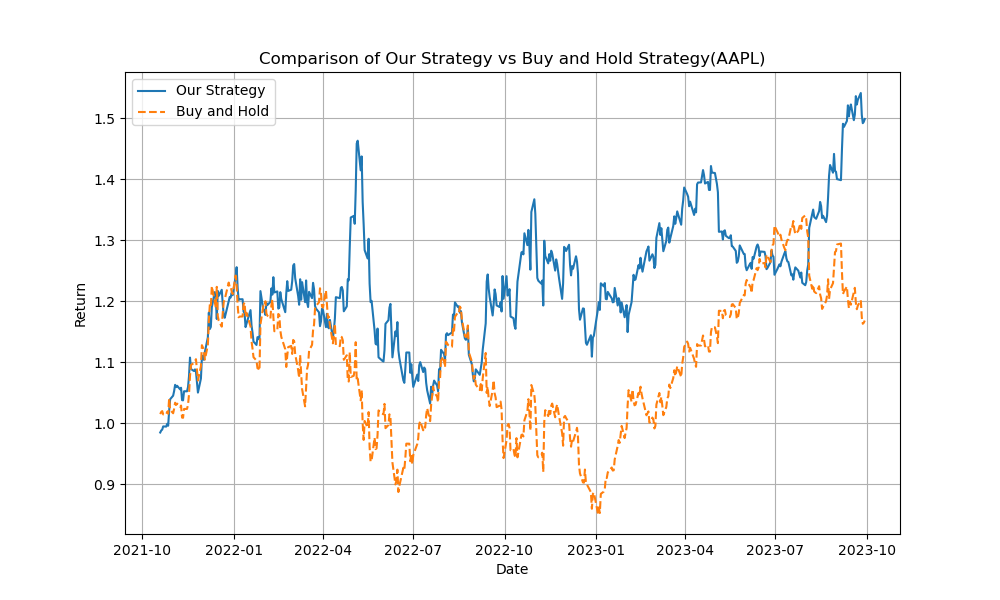}
        \caption{}
    \end{subfigure}
    \hfill
    \begin{subfigure}{0.5\textwidth}
        \includegraphics[width=\linewidth]{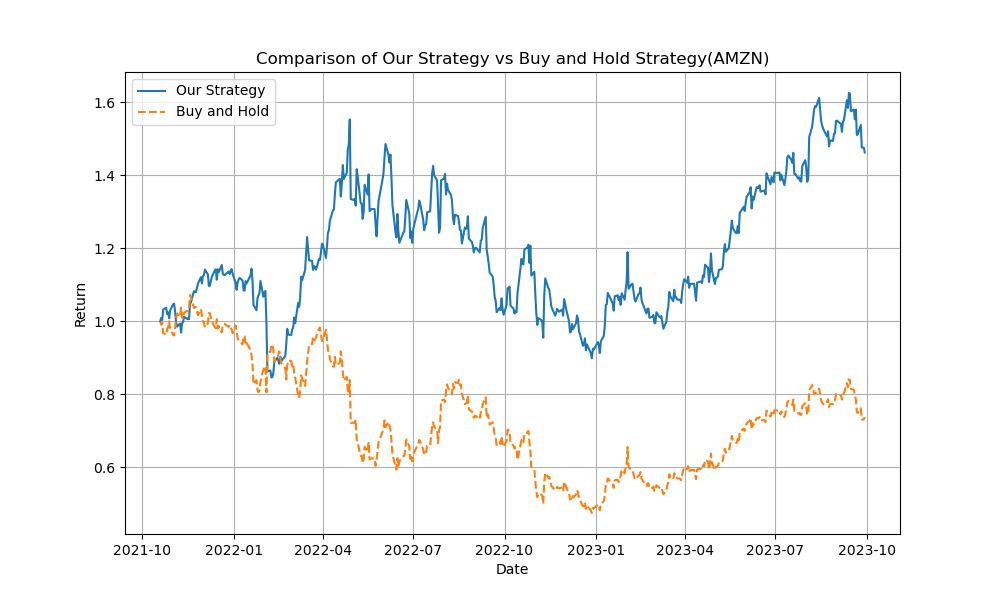}
        \caption{}
    \end{subfigure}
    \hfill
    \begin{subfigure}{0.5\textwidth}
        \includegraphics[width=\linewidth]{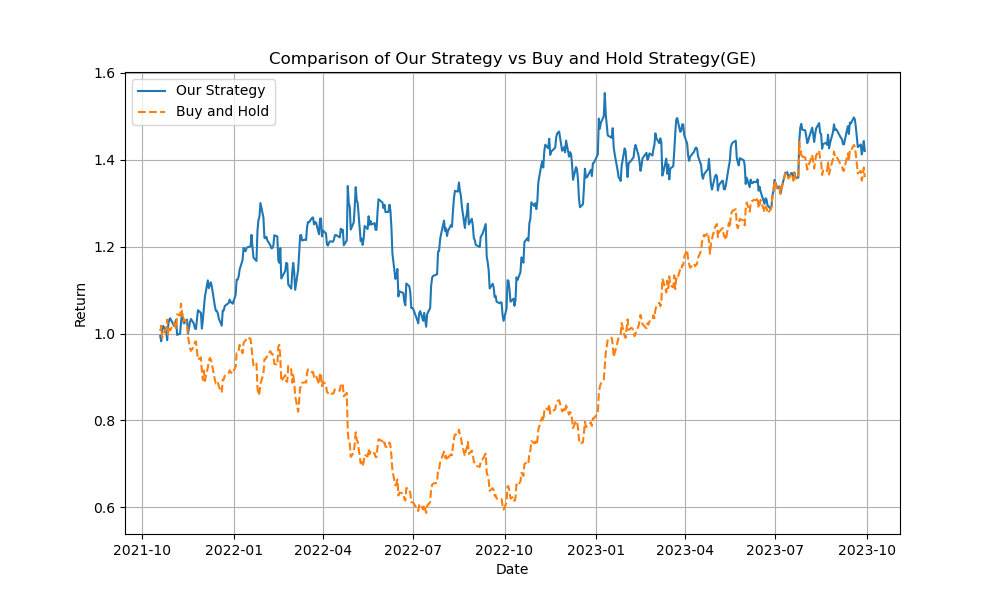}
        \caption{}
    \end{subfigure}
    \hfill
    \begin{subfigure}{0.5\textwidth}
        \includegraphics[width=\linewidth]{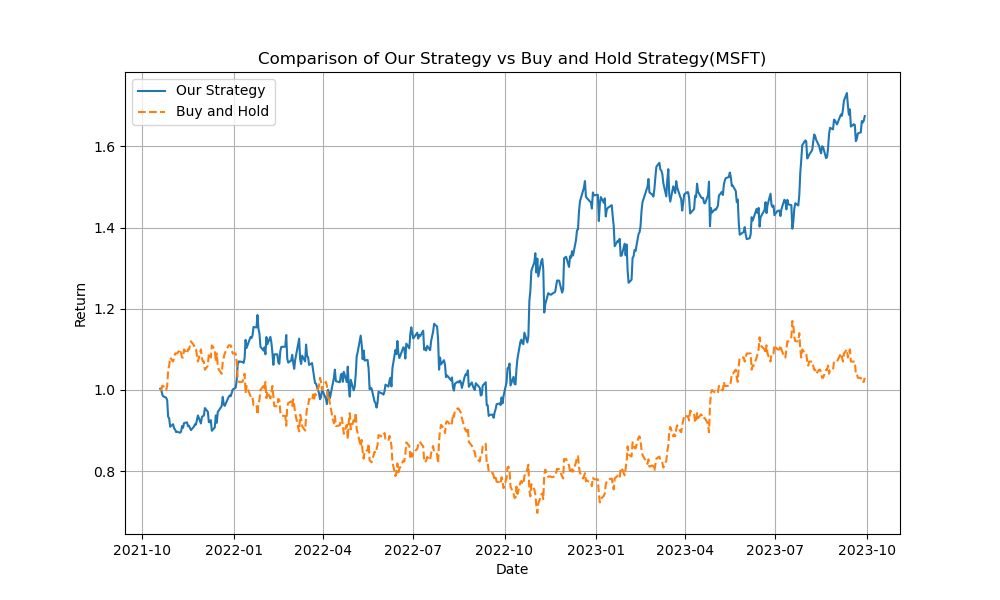}
        \caption{}
    \end{subfigure}
    \hfill
    \caption{Return of individual stock trading}
    \label{fig:multi_figs}
\end{figure}

Figures.3 presents the trading return results derived from our strategy against the returns from the buy and hold approach for each stock. Notably, during this post-pandemic period, large-cap stocks experienced significant drawdowns and elevated volatility under a buy and hold strategy. In contrast, the trading signals produced by our model consistently yielded desirable returns with lower maximum drawdowns, thereby demonstrating superior risk-adjusted performance.

\subsection{Portfolio Trading}

In addition to individual stock trading, we construct an equal-weighted portfolio, where each stock contributes equally to the overall investment on a daily basis. This portfolio trading strategy is designed to mitigate idiosyncratic risk and exploit diversification benefits.

\begin{figure}[!h]
    \centering
    \includegraphics[width=\textwidth]{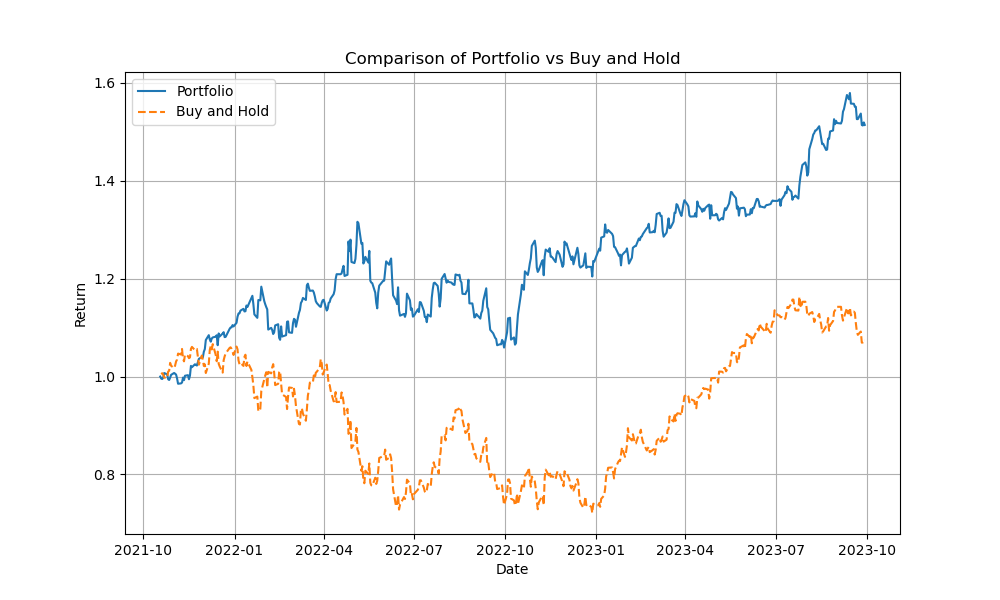}
    \caption{Total return of S\&P 500 stocks portfolio}
    \label{fig:single_figure}
\end{figure}

Figure.4 illustrates the cumulative return of the equal-weighted portfolio, alongside the return of a buy and hold portfolio. The results indicate that our portfolio trading strategy achieves a more robust return while maintaining a lower maximum drawdown compared to the individual stock trades. Conversely, the buy and hold portfolio exhibits only modest total returns paired with considerably larger drawdowns, emphasizing the advantages of dynamic signal-based portfolio management.

\subsection{Performance Evaluation}

A comprehensive performance evaluation is conducted to compare the effectiveness of individual stock trading, portfolio trading, and the conventional buy and hold strategy. Table 1 presents the key performance metrics, which include the annualized return, Sharpe ratio, and maximum drawdown for each trading strategy.

\begin{longtable}{@{}llccc@{}}
\caption{Trading Performance Metrics by Algorithm} \label{tab:forecast_metrics_by_algo}\\
\toprule
\textbf{Algorithm} & \textbf{Asset} & \textbf{Annualized Return} & \textbf{Sharpe Ratio} & \textbf{MDD} \\
\midrule
\endfirsthead

\multicolumn{5}{c}%
{\tablename\ \thetable\ -- \textit{Continued from previous page}} \\
\toprule
\textbf{Algorithm} & \textbf{Asset} & \textbf{Annualized Return} & \textbf{Sharpe Ratio} & \textbf{MDD} \\
\midrule
\endhead

\midrule
\multicolumn{5}{r}{\textit{Continued on next page}}\\
\endfoot

\bottomrule
\endlastfoot

\multirow{5}{*}{MLP}
    & AAPL & 11.02\% & 0.84  & 37.00\% \\
    & AMZN & 12.76\% & 0.93  & 31.92\% \\
    & GE   & 9.58\%  & 0.88  & 30.19\% \\
    & MSFT & 14.90\% & 1.01  & 31.21\% \\
    & Portfolio & 10.52\% & 0.95  & 29.31\% \\
    & B\&H Portfolio & 3.73\%  & -0.52 & 32.35\% \\
\midrule
\multirow{5}{*}{CNN}
    & AAPL & 12.51\% & 1.04 & 35.00\%  \\
    & AMZN & 13.82\% & 1.08 & 31.21\%  \\
    & GE   & 10.43\% & 1.01 & 29.41\%  \\
    & MSFT & 15.73\% & 1.12 & 30.54\%  \\
    & Portfolio & 11.72\% & 1.06 & 27.54\%  \\
    & B\&H Portfolio  & 3.73\%  & -0.52 & 32.35\%  \\
\midrule
\multirow{5}{*}{LSTM}
    & AAPL & 14.47\% & 1.18 & 34.11\%  \\
    & AMZN & 15.44\% & 1.23 & 30.02\%  \\
    & GE   & 10.94\% & 1.06 & 28.88\%  \\
    & MSFT & 17.27\% & 1.33 & 28.63\%  \\
    & Portfolio & 13.28\% & 1.21 & 25.14\%  \\
    & B\&H Portfolio & 3.73\%  & -0.52 & 32.35\%  \\
\midrule
\multirow{5}{*}{CNN-LSTM}
    & AAPL & 16.82\% & 1.33 & 30.24\%  \\
    & AMZN & 17.12\% & 1.38 & 28.92\%  \\
    & GE   & 12.51\% & 1.17 & 26.13\%  \\
    & MSFT & 20.17\% & 1.46 & 27.09\%  \\
    & Portfolio & 15.44\% & 1.31 & 23.04\%  \\
    & B\&H Portfolio & 3.73\%  & -0.52 & 32.35\%  \\
\midrule
\multirow{5}{*}{Attention-LSTM}
    & AAPL & 19.28\% & 1.67 & 25.43\%  \\
    & AMZN & 19.11\% & 1.63 & 25.17\%  \\
    & GE   & 15.73\% & 1.35 & 24.02\%  \\
    & MSFT & 23.41\% & 1.52 & 22.37\%  \\
    & Portfolio & 18.19\% & 1.47 & 20.83\%  \\
    & B\&H Portfolio & 3.73\%  & -0.52 & 32.35\%  \\
\midrule
\multirow{5}{*}{Our Method}
    & AAPL & 23.08\% & 2.21 & 18.78\%  \\
    & AMZN & 21.56\% & 1.94 & 18.21\%  \\
    & GE   & 19.75\% & 1.58 & 21.12\%  \\
    & MSFT & 30.32\% & 2.03 & 19.00\%  \\
    & Portfolio & 23.74\% & 1.84 & 14.01\%  \\
    & B\&H Portfolio & 3.73\%  & -0.52 & 32.35\%  \\

\end{longtable}
The data in Table 1 clearly indicate that our signal-based trading strategies yield superior performance compared to the buy and hold approach and other trading results generated by other algorithms, both in terms of return generation and risk management. The notably lower maximum drawdown and higher Sharpe ratios underscore the robustness and risk-adjusted benefits of our proposed methods.

\subsection{Evaluation Metrics}

To further assess the accuracy and robustness of our price prediction model, we employ a set of standard evaluation metrics, namely Mean Squared Error (MSE), Mean Absolute Error (MAE), Mean Absolute Percentage Error (MAPE), and the coefficient of determination (\(R^2\)). These metrics provide a quantitative basis for evaluating the predictive performance across different stocks and models.

\begin{longtable}{@{\extracolsep{\fill}} l l c c c c @{}}
\caption{Prediction Performance Metrics by Algorithm} \label{tab:forecast_metrics_by_algo}\\
\toprule
\textbf{Algorithm} & \textbf{Asset} & \textbf{MSE} & \textbf{MAE} & \textbf{MAPE} & \textbf{R Square} \\
\midrule
\endfirsthead

\multicolumn{6}{c}%
{\tablename\ \thetable\ -- \textit{Continued from previous page}} \\
\toprule
\textbf{Algorithm} & \textbf{Asset} & \textbf{MSE} & \textbf{MAE} & \textbf{MAPE} & \textbf{R Square} \\
\midrule
\endhead

\midrule
\multicolumn{6}{r}{\textit{Continued on next page}}\\
\endfoot

\bottomrule
\endlastfoot

\multirow{4}{*}{MLP} 
    & AAPL  & 4.5151 & 1.7239 & 0.0272 & 0.8331 \\
    & AMZN  & 1.9122 & 2.1143 & 0.0144 & 0.8212 \\
    & GE    & 3.1211 & 2.0895 & 0.0222 & 0.8411 \\
    & MSFT  & 2.6511 & 1.2559 & 0.0101 & 0.8184 \\
\midrule
\multirow{4}{*}{CNN} 
    & AAPL  & 4.2125 & 1.6932 & 0.0269 & 0.8901 \\
    & AMZN  & 1.7651 & 2.0012 & 0.0142 & 0.9102 \\
    & GE    & 2.8237 & 2.0199 & 0.0211 & 0.9032 \\
    & MSFT  & 2.3998 & 1.1922 & 0.0121 & 0.9231 \\
\midrule
\multirow{4}{*}{LSTM} 
    & AAPL  & 3.7421 & 1.6222 & 0.0263 & 0.9221 \\
    & AMZN  & 1.6623 & 1.9501 & 0.0137 & 0.9200 \\
    & GE    & 2.5651 & 1.9755 & 0.0208 & 0.9291 \\
    & MSFT  & 2.2143 & 1.1623 & 0.0134 & 0.9112 \\
\midrule
\multirow{4}{*}{CNN-LSTM} 
    & AAPL  & 3.6123 & 1.5723 & 0.0261 & 0.9456 \\
    & AMZN  & 1.5932 & 1.8332 & 0.0135 & 0.9236 \\
    & GE    & 2.4721 & 1.9227 & 0.0203 & 0.9357 \\
    & MSFT  & 2.1239 & 1.0912 & 0.0126 & 0.9491 \\
\midrule
\multirow{4}{*}{Attention-LSTM} 
    & AAPL  & 3.5482 & 1.5254 & 0.0260 & 0.9721 \\
    & AMZN  & 1.5401 & 1.7098 & 0.0134 & 0.9581 \\
    & GE    & 2.3923 & 1.8998 & 0.0201 & 0.9655 \\
    & MSFT  & 2.0697 & 1.0531 & 0.0110 & 0.9603 \\
\midrule
\multirow{4}{*}{Our Method} 
    & AAPL  & 3.5208 & 1.5031 & 0.0259 & 0.9879 \\
    & AMZN  & 1.5013 & 1.5987 & 0.0132 & 0.9801 \\
    & GE    & 2.3555 & 1.8734 & 0.0198 & 0.9855 \\
    & MSFT  & 2.0121 & 1.0169 & 0.0115 & 0.9887 \\

\end{longtable}

The evaluation metrics in Table 2 reflect a high level of predictive accuracy across all stocks, with \(R^2\) values exceeding 0.98 in most cases. These results confirm the model's efficacy in capturing complex market dynamics and further validate the reliability of the trading signals derived from the forecasts.

In summary, the experimental results demonstrate that our proposed trading strategy not only outperforms the traditional buy and hold approach in a high-volatility, post-pandemic market environment but also offers superior risk-adjusted returns both at the individual stock level and within a diversified portfolio. Other comparison algorithms performed steadily, but have limitation in dealing with high volatility market condition.

\section{Conclusion}
In this paper, we introduced a novel approach that integrates wavelet-transform convolution and channel attention mechanisms with an LSTM framework for stock price prediction and portfolio selection. A fixed set of stocks was employed to form the portfolio, and trading decisions were derived from the signals generated by the proposed predictive model. To facilitate a comprehensive comparison, five alternative models were also applied to the same task. The experimental evaluations encompassed price prediction accuracy, individual stock trading outcomes, portfolio-level trading strategies, and overall trading performance. The comparative analysis reveals that the proposed method effectively captures market dynamics, even under conditions of heightened volatility, thereby underscoring its robustness and practical potential in financial applications.

\bibliography{sn-bibliography}
\end{document}